\documentclass[10pt,onecolumn]{IEEEtran}

\IEEEoverridecommandlockouts

\usepackage{stmaryrd}
\usepackage{arydshln}
\usepackage{graphicx}
\usepackage{amsmath}
\usepackage{amssymb, verbatim}
\usepackage{mathrsfs}
\usepackage{dsfont}
\usepackage{url}
\usepackage[compress]{cite}
\usepackage{mdwlist}
\usepackage{paralist}
\usepackage{algorithmic}
\usepackage{algorithm}
\usepackage{tikz}
\usepackage{tkz-graph}
\usepackage[utf8]{inputenc}
\usepackage{multirow}
\usepackage{color}
\usepackage{todonotes}

\PassOptionsToPackage{usenames,dvipsnames,svgnames}{xcolor}
\usetikzlibrary{arrows,positioning,automata,calc}

\newtheorem{lemma}{\sc Lemma}
\newtheorem{theorem}[lemma]{\sc Theorem}
\newtheorem{definition}[lemma]{\sc Definition}
\newtheorem{proposition}[lemma]{\sc Proposition}
\newtheorem{corollary}[lemma]{\sc Corollary}
\newtheorem{remark}{\sc Remark}[section]
\newtheorem{assumption}{\sc Assumption}[section]
\newtheorem{example}{\sc Example}

\DeclareMathOperator*{\argmin}{arg\,min}

\renewcommand{\footnoterule}{%
  \kern -2pt
  \hrule width 0.3\textwidth height .5pt
  \kern 2pt
}

\definecolor{OliveGreen}{rgb}{0,0.5,0}

\newcommand{\N}{\llbracket n\rrbracket}
\newcommand{\C}[1]{\llbracket #1\rrbracket}
\newcommand{\V}{V}

\definecolor{lightgray}{rgb}{.80,.78,.69}

\begin{document}

\title{ Promoting Truthful Behaviour in Participatory-Sensing Mechanisms }
\author{
Farhad Farokhi, Iman Shames, Michael Cantoni\thanks{The authors are with the Department of Electrical and Electronic Engineering, the University of Melbourne, Parkville, Victoria 3010, Australia. This work is supported by a McKenzie Fellowship and the Australian Research Council (LP130100605).}}

\maketitle

\begin{abstract}
In this paper, the interplay between a class of nonlinear estimators and strategic sensors is studied in several participatory-sensing scenarios. It is shown that for the class of estimators, if the strategic sensors have access to noiseless measurements of the to-be-estimated-variable, truth-telling is an equilibrium of the game that models the interplay between the sensors and the estimator. Furthermore, performance of the proposed estimators is examined in the case that the strategic sensors form coalitions and in the presence of noise. 
\end{abstract}

\section{Introduction}
With a yearly expansion rate of 30\% for mobile broadband subscriptions and smartphones accounting for 65-70\% of all the sold mobile phones~\cite{EricssonConnected}, we are truly living in a connected world. This constant state of connectedness has enabled new technologies, such as participatory- and/or crowd-sensing applications, in which consented participants, with their smartphones, act as sensing units to estimate a variable\footnote{In this paper, we use the terms state, variable, and to-be-estimated-variable  interchangeably based on the context.}. Examples of commercial products using participatory-sensing schemes are Waze (for traffic estimation), Sensorly (for mobile coverage), Amazon review (for quality of service or product), or  Mobile Water Management (for  user data collection in control of irrigation canals) \cite{van2013canal,MobileTrack}. These systems, so far, have mainly relied on the benevolence of participants. However, due to various reasons, participants might provide false data. For instance, smartphones can be hacked or fake bots can be created by a hacker to feed false measurements to a sensing scheme~\cite{Waze2014}. Another reason could be that the individual participants might want to change the outcome of participatory-sensing schemes for their benefit. For instance, in crowd-sensing applications for traffic estimation, such as Waze, people inevitably realize that their reports change the traffic estimates which, in turn, diverts vehicles to and from their neighbourhoods~\cite{WazeLA2014}. Finally, it could also be that the participants wish to retain their privacy. For instance, people might provide inaccurate and misleading reports to a polling organization (one of the oldest forms of crowd-sensing applications) to avoid revealing private and/or sensitive information to governments or for-profit organizations. Therefore, we are interested in designing robust participatory-sensing schemes that can work reliably in the face of strategic false-data injection.

An earlier study in estimation with strategic sensors showed that, counter-intuitively, the performance of linear estimators degrades as the number of sensors increases~\cite{6859123}. Further, it was shown that when the sensors are herding, i.e., they are imitating each other's policies, the quality of the estimation improves with the summoning of more sensors. Herding behaviour could be caused by the bounded rationality of the sensors. However, it can be also induced by designing an appropriate estimator that pushes each sensor to ``behave the same as the rest''. Here, to utilize this observation, we design nonlinear estimators that can reject outlier reports and, hence, induce a herding behaviour among the sensors. A median estimator is an example of such an estimator. The fact that median is robust in the presence of noisy and corrupted data is well known, e.g.~see \cite{hoaglin1983understanding}. However, to the best of our knowledge, the benefits of using a median estimator in the presence of strategic sensors and in a game-theoretic framework has not been formalized in the literature.

The main contributions of the paper are as follows. First, two linear estimators are considered and it is shown that sensors engaging in a  \emph{truth-telling} behaviour, i.e.~reporting the correct variable that the participatory-sensing scheme wishes to estimate, does not correspond to an equilibrium of the game modelling the interaction between the sensors and the estimator. Next, it is demonstrated that for a class of nonlinear estimators  truth-telling is in fact an equilibrium of this game. Third, the scenarios where strategic sensors might be colluding and form coalitions is studied. Particularly, the link between the size of the coalitions and the performance of the estimators in the aforementioned class is established. Fourth, the case where the measurements carried out by the sensors are noisy is studied.  

Many participatory-sensing mechanisms adopt quality assessment procedures and provide appropriate incentives to extract useful data. In this paper, however, we show that even if the participatory-sensing services do not adopt quality-assessment procedures or provide incentives, they can shape the behaviour of strategic sensors to elicit a truthful message. This is certainly a favourable approach as (\textit{i}) there is no need for incentives (in either monetary or psychological forms) in the often large groups of recruited participants and (\textit{ii}) it further reduces the complexity of the employed estimators.

The rest of the paper is as follows. In Section~\ref{sec:noiseless}, we investigate the interplay between the strategic sensors and the linear as well as nonlinear estimators. Section~\ref{sec:coalition} extends these results to the case where the sensors can form coalitions. In Section~\ref{sec:noisy}, we study nonlinear estimators when the sensors are noisy. Finally, we conclude the paper in Section~\ref{sec:conc}.

\section{Noiseless Measurements} \label{sec:noiseless}
Let us consider the case where a receiver is interested in measuring the variable $x\in\mathbb{R}$. Hence, it employs $n>1$ sensors to measure this variable and report it back. Each sensor transmits a measurement $y_i\in\mathbb{R}$. We assume that the transmissions occur simultaneously and, thus, the sensors do not have access to the messages communicated by each other. The receiver subsequently uses these messages to construct an estimate of $x$, denoted by $\hat{x}\in\mathbb{R}$. The receiver wants to guarantee that the estimation error $\mathbf{E}\{\|x-\hat{x}\|_2\}$ is small, where $\mathbf{E}\{\cdot\}$ denotes the expectation of its argument. In this section, we assume that the sensors have access to the noiseless measurements of $x$. However, their interests are not aligned with that of the receiver and each other, that is, sensor $i\in\N$ wishes to minimize the cost $\mathbf{E}\{\|(x+\theta_i)-\hat{x}\|_2\}$, where $\theta_i\in\mathbb{R}$ is its private information (i.e., it is not known by the other sensors and the receiver). Here, $\N$ denotes the set $\{1,\dots,n\}$.

\begin{assumption} $x$ and $(\theta_i)_{i\in\N}$ are jointly distributed Gaussian random variables with zero mean.
\end{assumption}

We are interested in large groups of homogeneous sensors to mimic the behaviour of large crowds. Therefore, we make the following assumption.

\begin{assumption} $(\theta_i)_{i\in\N}$ are identically and independently distributed random variables. Moreover, $x$ and $\theta_i$ are independent for all $i\in\N$.
\end{assumption}
%

Let $g_i(\cdot |x,\theta_i)$ denote the conditional distribution that sensor~$i$ uses for generating its message $y_i$ (i.e., its policy). Therefore, for any Lebesgue-measurable set $\mathcal{Y}_i\subseteq \mathbb{R}$, we get
\begin{align*}
\mathbf{P}\{y_i\in\mathcal{Y}_i\}=\int_{\xi\in\mathcal{Y}_i} g_i(\xi|x,\theta_i)\mathrm{d}\xi.
\end{align*}
We use the notation $\mathcal{G}_i$ to denote the set of all such conditional distributions for each sensor $i$. Moreover, since the dimension of the messages for all the sensors is the same, $\mathcal{G}_1=\dots=\mathcal{G}_n=\mathcal{G}$. Let us define a cost for sensor $i\in\N$ as
\begin{align*}
V_i((g_i)_{i\in\N};\pi)
&= \mathbf{E}\{\|(x+\theta_i)-\pi((y_i)_{i\in\N})\|_2\} \\
&=\int_{-\infty}^{\infty}\cdots \int_{-\infty}^{\infty}\|(x+\theta_i)-\pi((y_i)_{i\in\N})\|_2
\prod_{j\in\N}[g_j(y_j|x,\theta_j)\mathrm{d}y_j] \prod_{j\in\N}[p_{\theta}(\theta_j)\mathrm{d}\theta_j] p_x(x)\mathrm{d}x,
\end{align*}
where $(g_i)_{i\in\N}\in\mathcal{G}^n$ is the policy of all the sensors, $\pi:\mathbb{R}^n\rightarrow\mathbb{R}$ is the policy of the receiver, i.e., $\hat{x}=\pi((y_i)_{i\in\N})$, and $p_x$ and $p_{\theta}$ are probability density functions. Note that this is an \textit{ex ante} cost function as the sensors do not wait until they receive their private information and the state measurement in forming the cost function (it leads to a setup in which the sensors select their policies before entering the game). Hence, at the equilibrium induced by this cost function, the parameters of the policy are not a function of the private information and the state measurement, however, the transmitted message can be a function of these measurements\footnote{Using an \textit{ex ante} optimal policy can be motivated by the lack of enough computational resources for online calculation of the policy based on the realization of the private information and the state. }. Alternatively, we can form an \textit{ex post} cost function
\begin{align*}
U_i((g_i)_{i\in\N};\pi)
&= \mathbf{E}\{\|(x+\theta_i)-\pi((y_i)_{i\in\N})\|_2|\theta_i,x\} \\
&=\int_{-\infty}^{\infty}\cdots \int_{-\infty}^{\infty}\|(x+\theta_i)-\pi((y_i)_{i\in\N})\|_2
\prod_{j\in\N} [g_j(y_j|x,\theta_j)\mathrm{d}y_j] \prod_{j\in\N\setminus\{i\}}^n [p_{\theta}(\theta_j)\mathrm{d}\theta_j] .
\end{align*}

\begin{definition}[$\pi$-Stackelberg Equilibrium] A tuple of conditional distributions $(g_i^*)_{i\in\N}\in\mathcal{G}^n$ constitutes an \textit{ex~ante} $\pi$-Stackelberg equilibrium if
\begin{align*}
g_j^*\in\argmin_{g_j\in\mathcal{G}} V_j(g_j,(g^*_i)_{i\in\N\setminus\{j\}};\pi), \,\forall j\in\N.
\end{align*}
The tuple constitutes an \textit{ex post} $\pi$-Stackelberg equilibrium if
\begin{align*}
g_j^*\in\argmin_{g_j\in\mathcal{G}} U_j(g_j,(g^*_i)_{i\in\N\setminus\{j\}};\pi), \,\forall j\in\N.
\end{align*}
In statements where we do not distinguish between \textit{ex ante} and \textit{ex post} equilibria, the statement holds in both senses.
\end{definition}

Note that $\pi$, in $\pi$-Stackelberg equilibrium, is a generic place-holder for an arbitrary policy $\pi$ and indicates that the equilibrium corresponds to this policy.

In this paper, our interest is to find an estimator $\pi$ that can extract useful information from strategic sensors. This is motivated by the observation that, for linear estimators, the quality of the estimation degrades as the number of participating sensors increases~\cite{6859123}. Let us start with a simple, yet widely used, linear estimator to illustrate the problem.

\begin{definition}[Averaging Estimator] The receiver employs the estimator
\begin{align*}
\hat{x}=\psi((y_i)_{i\in\N}):=(y_1+\dots+y_n)/n.
\end{align*}
A tuple of conditional distributions $((g_i^*)_{i\in\N})\in\mathcal{G}^n$ constitutes an \textit{ex ante} (\textit{ex post}) equilibrium for the averaging estimator if it is an \textit{ex ante} (\textit{ex post}) $\psi$-Stackelberg equilibrium.
\end{definition}

\begin{definition}[Truth-Telling Portfolio] Sensor $i$ follows the truth-telling strategy if\footnote{This is the same as saying $\mathbf{P}\{y_i=x\}=1$.} $g_i(y_i|x,\theta_i)=\delta(y_i-x)$, where $\delta$ is the Dirac delta distribution\footnote{The Dirac delta distribution is a mapping $\delta:\mathbb{R}\rightarrow \mathbb{R}\cup\{\pm\infty\}$ such that $\delta(t)=0$ for all $t\in\mathbb{R}\setminus\{0\}$ and $\int_{-\infty}^\infty \delta(t)\mathrm{d}t=1$.}. The truth-telling portfolio is a tuple of conditional distributions such that all the sensors are employing the truth-telling strategy.
\end{definition}

Now, we can prove the following negative result regarding the averaging estimator.

\begin{theorem} \label{tho:average} The truth-telling portfolio is not an equilibrium for the averaging estimator.
\end{theorem}

\begin{IEEEproof} Let all the players except player~$i\in\N$ employ the truth-telling strategy. Therefore, $y_j=x$ for all $j\in\N\setminus\{i\}$. Hence, we have $\hat{x}=(1-1/n)x+(1/n)y_i.$
Now, sensor $i$ using the policy $y_i=x+n\theta_i$ results in $\mathbf{E}\{\|(x+\theta_i)-\hat{x}\|_2|\theta_i,x\}=0$ and $\mathbf{E}\{\|(x+\theta_i)-\hat{x}\|_2\}=0$, which is strictly less than, respectively, substitution of the truth-telling strategy in both \textit{ex~post} and \textit{ex~ante} cost functions.\end{IEEEproof}

Even with the optimal linear estimator $\mathbf{E}\{x\,|\,y_1,\dots,y_n\}$, it was observed in~\cite{6859123} that the truth-telling is not an equilibrium of the game and that, at the equilibrium, the quality of the estimation degrades as the number of sensors increases. In limit, no information can be recovered from the transmitted messages. This observation, together with Theorem~\ref{tho:average}, motivates us to find estimators for which the truth-telling portfolio is an equilibrium.  We address this concern in the remainder of this section.

\begin{definition}[$2\ell$-Rejection Averaging Estimator] \label{def:rejavestimator} Assume that $n\geq 2\ell+1$. Let $(i_j)_{j\in\N}$ be given such that $y_{i_1}\leq y_{i_2} \leq \cdots \leq y_{i_n}.$
The receiver employs the estimator
\begin{align*}
\hat{x}=\psi_\ell((y_i)_{i\in\N}):=\frac{1}{n-2\ell}\sum_{j=\ell+1}^{n-\ell} y_{i_j}.
\end{align*}
A tuple of conditional distributions $((g_i^*)_{i\in\N})\in\mathcal{G}^n$ constitutes an \textit{ex ante} (\textit{ex post}) equilibrium for the $2\ell$-rejection averaging estimator if it is an \textit{ex ante} (\textit{ex post}) $\psi_\ell$-Stackelberg equilibrium.
\end{definition}

\begin{remark} In the statistics literature (e.g.,~\cite[p.\,16]{davison2003statistical}), $2\ell$-rejection averaging estimators are alternatively known as trimmed averaging estimators (since they are derived from an averaging estimator by excluding the extreme values). Here, we use the name $2\ell$-rejection averaging estimator because of our desire to work with integer values of $\ell$ rather than percentages of rejection $\ell/n$.  
\end{remark}

\begin{theorem} \label{tho:2lreject} The truth-telling is an equilibrium for the $2\ell$-rejection averaging estimator for all $\ell\in\C{\lfloor(n-1)/2\rfloor}$.
\end{theorem}

\begin{IEEEproof} Let all the players except player~$i\in\N$ employ the truth-telling strategy. Therefore, $y_j=x, \forall j\in\N\setminus\{i\}$. Hence, we have $\hat{x}=x$ irrespective of $y_i$ (as it will be rejected). Thus, truth-telling (among all the other policies) minimizes the cost of sensor~$i$ in both \textit{ex ante} and \textit{ex post} senses.
\end{IEEEproof}

\begin{definition}[Median Estimator] Let $(i_j)_{j\in\N}$ be given such that
$y_{i_1}\leq y_{i_2} \leq \cdots \leq y_{i_n}.$
The receiver employs the estimator
\begin{align*}
\hat{x}=\phi((y_i)_{i\in\N}):=
\begin{cases}
(y_{i_{n/2}}+y_{i_{n/2+1}})/2, & n\in\mathbb{E},\\
y_{i_{(n+1)/2}}, & n\in\mathbb{O},
\end{cases}
\end{align*}
where $\mathbb{E}$ and $\mathbb{O}$ represent the sets of even and odd integers, respectively.
A tuple of conditional distributions $((g_i^*)_{i\in\N})\in\mathcal{G}^n$ constitutes an \textit{ex ante} (\textit{ex post}) equilibrium for the median estimator if it is an \textit{ex ante} (\textit{ex post}) $\phi$-Stackelberg equilibrium.
\end{definition}

\begin{lemma} The median estimator is equivalent to the $2\ell$-rejection averaging estimator if $\ell=(n-2)/2$ for $n\in\mathbb{E}$ and $\ell=(n-1)/2$ for $n\in\mathbb{O}$.
\end{lemma}

\begin{IEEEproof} The proof follows from simple algebraic manipulations and is hence omitted.
\end{IEEEproof}

\begin{corollary} \label{cor:median} The truth-telling is an equilibrium for the median estimator.
\end{corollary}

So far, we have assumed that the sensors do not form coalitions to deceive the receiver. In the next section, we define a different game in which sensors can act together. 

\section{Extension to Coalitions} \label{sec:coalition}
Assume that sensor $i\in\N$ can submit $c_i\in \mathbb{N}$ messages instead of one. This setup has two interpretations. First, each sensor represents a coalition of size $c_i$ instead of a single sensor. Alternatively, we can assume that each sensor represents an array of sensors introduced by a single hacker. Therefore, the receiver has access to $(y_i)_{i\in\C{c}}$ where $c=\sum_{j\in\N}c_j$. The receiver does not know $(c_j)_{j\in\N}$. Here, the definition of the $2\ell$-rejection averaging estimator  is the same as in Definition~\ref{def:rejavestimator} with $c$ denoting the number of messages instead of $n$.

Examples of participatory-sensing services that admit coalitions are legislative bodies, e.g.~the U.S. congress, and truth-finding committees, e.g.~royal commissions in most of the Commonwealth countries. Here, the coalitions are political parties because their members most often, persuaded by the party whip,  vote on party lines. Therefore, it would be nice to construct estimators that can recover the truth despite the ever-growing partisanship.

\begin{theorem} \label{tho:coalition} Let $\sum_{j\in\N\setminus\{i\}}c_j \geq c_i+1$ for all $i\in\N$. The truth-telling is an equilibrium for the $2\ell$-rejection averaging estimator if $\max_{i\in\N}c_i\leq \ell\leq \lfloor (\sum_{j\in\N}c_j-1)/2 \rfloor$.
\end{theorem}

\begin{IEEEproof} The proof is similar to the proof of Theorem~\ref{tho:2lreject} and is hence omitted.
\end{IEEEproof}

\begin{remark} Theorem~\ref{tho:coalition} shows that, for a given $\ell$, the estimator is robust to admitting a coalition of sensors with the size of, at most, $\ell$ assuming that no coalition has more sensors than the sum of the size of all other coalitions minus one (i.e., there is a balance of power between the competing coalitions).
\end{remark}

\begin{corollary} \label{cor:coalition} Let 
\begin{align*}
\min_{i\in\N} \left[\left(\sum_{j\in\N\setminus\{i\}} c_j \right)-c_i\right]\geq
\begin{cases}
1, & \sum_{j\in\N} c_j\in\mathbb{O}, \\
2, & \sum_{j\in\N} c_j\in\mathbb{E}.
\end{cases}
\end{align*}
The truth-telling portfolio is an equilibrium for the median estimator.
\end{corollary}

\begin{remark} Corollary~\ref{cor:coalition} shows that the truth-telling portfolio is an equilibrium if no individual coalition has the majority. Thus, the median estimator is extremely robust to manipulation by strategic entities even if they cooperate. This observation has interesting implications in politics, that is, any truth finding committee, as a whole, can only recover the truth so long as no single party has the majority because, in such case, they can silence the voice of the others.
\end{remark}

\section{Extension to Noisy Measurements} \label{sec:noisy}
Consider the case where sensor~$i\in\N$ has access to noisy measurements of the state denoted by $z_i=x+w_i$, where $(w_i)_{i\in\N}$ are independent zero-mean Gaussian random variables. Similarly, sensor $i$ uses the conditional distribution $g_i(\cdot |z_i,\theta_i)$ to generate its message $y_i$. In this case, we say that sensor $i$ follows the truth-telling strategy if $g_i(y_i|z_i,\theta_i)=\delta(y_i-z_i)$. Unfortunately, access to noisy measurements destroys the truth-telling property of the median estimator.

\begin{theorem} The truth-telling portfolio is not an equilibrium for the $2\ell$-rejection estimator in the presence of noise.
\end{theorem}

\begin{IEEEproof} Let us pick a sensor $k\in\N$. Set $y_j=z_j$ for all $j\in\N\setminus\{k\}$. Assume that $y_k=(n-2\ell)\theta_k+z_k$. Let $(i_j)_{j=1}^{n-1}$ be given such that $i_j\neq k$ for all $1\leq j\leq n-1$ and $y_{i_1}\leq y_{i_2} \leq \cdots \leq y_{i_{n-1}}.$ Note that $k\in\{\ell+1,\dots,n-\ell\}$ with a positive probability. In that case, we have $\hat{x}=1/(n-2\ell)\sum_{j=\ell+1}^{n-\ell} y_{i_j}.$  This results in a cost equal to $\mathbb{E}\{\|(x+\theta_k)-\hat{x}\|_2|z_k,\theta_k,y_1,\dots,y_n\}=\mathbb{E}\{\|x-1/(n-2\ell)\sum_{j=\ell+1}^{n-\ell} z_{i_j}\|_2|z_k,\theta_k,y_1,\dots,y_n\}$ which is strictly smaller than sensor $k$'s cost had it been truthful: $\mathbb{E}\{\|(x+\theta_k)-1/(n-2\ell)\sum_{j=\ell+1}^{n-\ell} z_{i_j}\|_2|z_k,\theta_k,y_1,\dots,y_n\}$. Hence, by taking expectation of these terms over $y_1,\dots,y_n$, we can show that the cost of sensor $k$ can be reduced by not acting truthfully. This concludes the proof.
\end{IEEEproof}

A similar result can be proved for the median estimator.

\begin{theorem} \label{tho:median:noisy} The truth-telling portfolio is not an equilibrium for the median estimator in the presence of noise.
\end{theorem}

\begin{IEEEproof}  Let $n\in\mathbb{O}$ as, with a similar idea, we can prove the result for $n\in\mathbb{E}$. Pick a sensor $k\in\N$. Let $(i_j)_{j=1}^{n-1}$ be given such that $i_j\neq k$ for all $1\leq j\leq n-1$ and $y_{i_1}\leq y_{i_2} \leq \cdots \leq y_{i_{n-1}}.$
Hence, we have
\begin{align*}
\hat{x}=
\begin{cases}
y_{i_{(n-1)/2}}, & y_k<y_{i_{(n-1)/2}},\\
y_k, & y_{i_{(n-1)/2}}\leq y_k\leq y_{i_{(n+1)/2}},\\
y_{i_{(n+1)/2}}, & y_k>y_{i_{(n+1)/2}},
\end{cases}
\end{align*}
This gives
\begin{align*}
\mathbf{E}&\{\|(x+\theta_k)-\hat{x}\|_2|y_1,\dots,y_n\}=
\begin{cases}
\mathbf{E}\{\|(x+\theta_k)-y_{i_{(n-1)/2}}\|_2\}, & y_k<y_{i_{(n-1)/2}},\\
\mathbf{E}\{\|(x+\theta_k)-y_{k}\|_2\}, & y_{i_{(n-1)/2}}\leq y_k\leq y_{i_{(n+1)/2}},\\
\mathbf{E}\{\|(x+\theta_k)-y_{i_{(n+1)/2}}\|_2\}, & y_k>y_{i_{(n+1)/2}}.
\end{cases}
\end{align*}
Selecting $y_k=\mathbf{E}\{x+\theta_k|\theta_k,z_k\}=z_k+\theta_k$ minimizes $\mathbf{E}\{\|(x+\theta_k)-y_{k}\|_2\}$. This results in a strictly smaller cost than using a truthful strategy since $y_{i_{(n-1)/2}}\leq z_k+\theta_k\leq y_{i_{(n+1)/2}}$ occurs with a positive probability. This concludes the proof.
\end{IEEEproof}

Although truth-telling is no longer an equilibrium, we can characterize another equilibrium that can reveal some information about the to-be-estimated-variable.

\begin{remark} Notice that, without loss of generality, we can consider an odd number of measurements because we can always transform an even number of measurements into an odd number by either discarding a measurement randomly or by introducing a very large or a very small measurement (that always gets discarded). \end{remark}

\begin{theorem} \label{tho:median:noisy:2} Let $n\in\mathbb{O}$. The tuple $(g_i^*)_{i\in\N}\in\mathcal{G}^n$ defined as $g_i^*(y_i|z_i,\theta_i)=\delta(y_i-(z_i+\theta_i))$, $i\in\N$, is an equilibrium for the median estimator in the presence of noise.
\end{theorem}

\begin{IEEEproof} Following the same line of reasoning as in the proof of Theorem~\ref{tho:median:noisy}, selecting $y_k=z_k+\theta_k$ is the best response of each sensor irrespective of the others in both \textit{ex ante} and \textit{ex post} senses.
\end{IEEEproof}

\begin{remark} It is interesting to note that reporting $y_k=z_k+\theta_k$ is a dominant strategy, i.e., it is in the benefit of players irrespective of other reports (when the players do not form collations). Therefore, even if some sensors are randomly-behaving or faulty, the rational ones report $z_k+\theta_k$.
\end{remark}

\begin{remark} The captured equilibrium in Theorem~\ref{tho:median:noisy:2} is not unique. This can be observed from the fact that  the sensors can employ any stochastic or deterministic mappings for constructing their messages $y_k$ when it is very large or very small, since the message  will be discarded regardless and has no impact on the outcome of the estimation.
\end{remark}

\begin{proposition} The equilibrium in Theorem~\ref{tho:median:noisy:2} results in $\lim_{m\rightarrow \infty,n=2m+1}\mathbf{E}\{\|x-\phi((y_i)_{i\in\N})\|_2\}=0.$
\end{proposition}

\begin{IEEEproof} At the equilibrium, we have $y_k=z_k+\theta_k=x+w_k+\theta_k$. Therefore, $\hat{x}=\phi((x+w_i+\theta_i)_{i\in\N})=x+\phi((w_i+\theta_i)_{i\in\N}).$
From~\cite{Biometrika1931}, we know that $\lim_{m\rightarrow \infty,n=2m+1}\phi((w_i+\theta_i)_{i\in\N})=0, \mathrm{a.s.}$
This concludes the proof.
\end{IEEEproof}

\begin{figure}
\centering
\includegraphics[width=0.5\linewidth]{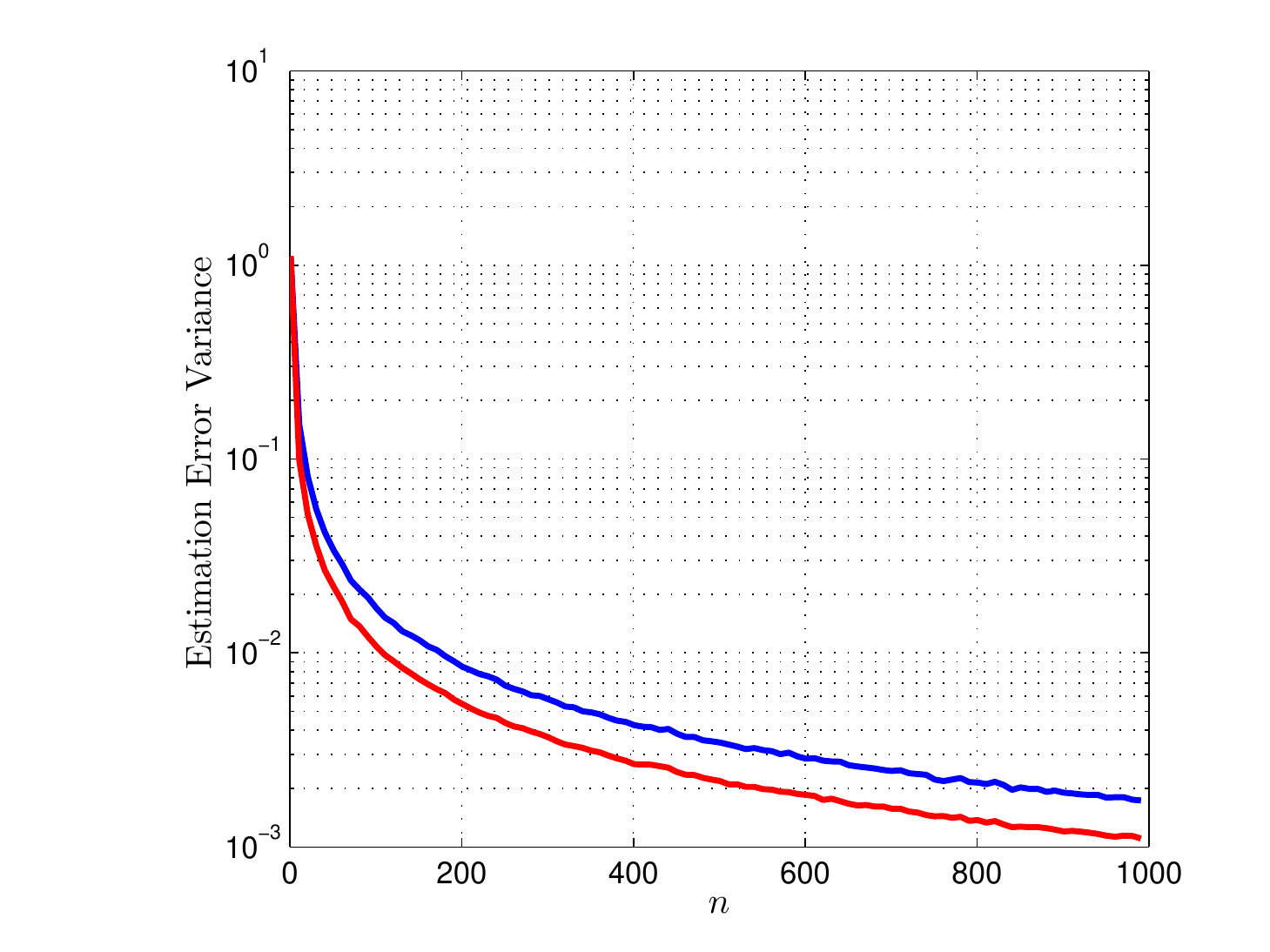}
\caption{ \label{fig:1} Estimation error variance for median (blue) and mean (red) estimators at the equilibrium captured in Theorem~\ref{tho:median:noisy:2}. }
\end{figure}

\begin{example}
Consider an estimation problem in which $V_{xx}=1$, $\V_{\theta_i\theta_i}=1$, and $V_{w_iw_i}=0.1$ for all $i$ where $V_{xx}$, $\V_{\theta_i\theta_i}$, and $V_{w_i w_i}$ are the variances of $x$, $\theta_i$, and $w_i$, respectively. When using the median estimator, at the equilibrium in Theorem~\ref{tho:median:noisy:2}, the sensors report $y_i=z_i+\theta_i, \forall i$. The blue curve in Fig.~\ref{fig:1} shows the estimation error $\mathbf{E}\{\|x-\phi((y_i)_{i\in\N})\|_2\}$ approximated using 10000 random samples. If, for the signals transmitted at this specific equilibrium, we were to use the averaging estimator, the estimation error would follow the red curve in Fig.~\ref{fig:1}, which is certainly smaller. This degradation in the performance is the price of robustness. One way to capitalize on this difference is to lie to the players that the utilized estimator is the median one but use an averaging policy and, hence, \emph{nudge} them towards a good behaviour (from the perspective of the estimator)~\cite{thaler2009nudge}. Note that this is applicable if the players cannot infer the correct mechanism by experimenting, e.g., when the players interact with a given participatory-sensing scheme very infrequently.\hfill $\lozenge$
\end{example}

\section{Conclusions} \label{sec:conc}
In this paper, the problem of designing participatory-sensing mechanisms is considered. Particularly, it is shown that for a class of nonlinear estimators, truth telling is an equilibrium of game modelling the interaction between the sensors and the estimator. Later, it is established, for the case where the sensors collude and form coalitions that are no lager than half of the total participants, that there is always an estimator, specifically the median estimator, which results in truth telling being an equilibrium.  Future research can focus on dynamic estimation problems. 

\bibliographystyle{ieeetr}
\bibliography{compile_new}

\begin{thebibliography}{10}

\bibitem{EricssonConnected}
{Ericsson AB}, ``Ericsson mobility report.'' www.ericsson.com, 2014.
\newblock Online; posted November 2014;
  \url{http://www.ericsson.com/res/docs/2014/ericsson-mobility-report-november-2014.pdf}.

\bibitem{van2013canal}
P.-J.~V. Overloop, ``Canal control system,'' July~4 2013.
\newblock WO Patent App. PCT/NL2012/050,893,
  http://www.mobilewatermanagement.com/.

\bibitem{MobileTrack}
P.~van Overloop, J.~Davids, and M.~M. Vierstra, ``Mobile monitoring
  technologies: The mobiletracker and the remotetracker,'' in {\em USCID
  Conference}, (Sacramento, CA), 2014.

\bibitem{Waze2014}
N.~Tufnell, ``Students hack {Waze}, send in army of traffic bots.''
  wired.co.uk, 2014.
\newblock Online; posted 25 March 2014;
  \url{http://www.wired.co.uk/news/archive/2014-03/25/waze-hacked-fake-traffic-jam}.

\bibitem{WazeLA2014}
{Daily Mail}, ``Residents outrage after {Waze} app used to avoid traffic ends
  up sending {Los Angeles} drivers down once quiet `hidden' street.'' Daily
  Mail, 2014.
\newblock Online; posted 15 December 2014;
  \url{http://www.dailymail.co.uk/news/article-2873468/People-finding-waze-hidden-streets.html}.

\bibitem{6859123}
F.~Farokhi, A.~M.~H. Teixeira, and C.~Langbort, ``Gaussian cheap talk game with
  quadratic cost functions: When herding between strategic senders is a
  virtue,'' in {\em Proceedings of the American Control Conference},
  pp.~2267--2272, 2014.

\bibitem{hoaglin1983understanding}
D.~C. Hoaglin, F.~Mosteller, and J.~W. Tukey, {\em Understanding robust and
  exploratory data analysis}, vol.~3.
\newblock Wiley New York, 1983.

\bibitem{davison2003statistical}
A.~C. Davison, {\em Statistical Models}.
\newblock Cambridge Series in Statistical and Probabilistic Mathematics,
  Cambridge University Press, 2003.

\bibitem{Biometrika1931}
T.~Hojo and K.~Pearson, ``Distribution of the median, quartiles and
  interquartile distance in samples from a normal population,'' {\em
  Biometrika}, vol.~23, no.~3/4, pp.~315--363, 1931.

\bibitem{thaler2009nudge}
R.~H. Thaler and C.~R. Sunstein, {\em Nudge: Improving Decisions about Health,
  Wealth, and Happiness}.
\newblock Penguin Group US, 2009.

\end{thebibliography}

\end{document}